**Title:**

# Local intercellular coupling is sufficient for long-range calcium signaling

**Running Title:** Local coupling and calcium signaling

**Authors and Affiliations:**

**Benjamin M. Goykadosh[1],** Vasuretha Chandar[1], Harikrishnan Parameswaran[1*]

[1]Department of Bioengineering, Northeastern University, Boston, MA, USA

Corresponding Author* Email: h.parameswaran@northeastern.edu

## Abstract

Long-range intercellular calcium ($Ca^{2+}$) signaling coordinates biological processes ranging from fertilization to contraction and cell death. The classical model attributes this long-range propagation to rapid diffusion of inositol 1,4,5-trisphosphate ($IP_3$) through gap junctions. However, recent evidence that $IP_3$ diffuses far more slowly than previously believed, and that $Ca^{2+}$ oscillations persist even when gap junctions are disassembled, indicates that an alternative mechanism must sustain long-range communication. Here we develop a computational model showing that local coupling between neighboring cells is sufficient to generate and propagate regenerative $Ca^{2+}$ oscillations across a cell population without fast molecular diffusion. Each cell is treated as an oscillator whose intrinsic frequency is set by its local $IP_3$ concentration through an $IP_3$-dependent refractory period, and neighboring cells are coupled using a Kuramoto nearest-neighbor framework. In a dual-stiffness regime, cells on a stiff extracellular matrix entrain their soft-matrix neighbors, producing an offset traveling wave of $Ca^{2+}$ release. This reproduces the finite spatial range of influence (~8 cell lengths) observed experimentally. Our findings propose a diffusion-independent paradigm for calcium signaling in which local intercellular coupling drives long-range communication, offering insight into how localized ECM stiffening in asthma and fibrosis may produce systemic effects.

## Statement of Significance

Cells coordinate their behavior across long distances through waves of calcium, a process central to functions ranging from insulin production to wound healing. One prevailing explanation holds that these long-range signals depend on the rapid diffusion of a small molecule, $IP_3$, between cells. However, $IP_3$ diffuses too slowly to account for this coordination, leaving the mechanism unclear. Using a computational model, we show that long-range calcium signaling can instead emerge from purely local interactions between neighboring cells. Our results offer a new framework for understanding how cells communicate over long distances and suggest how localized stiffening of tissue, as occurs in asthma and fibrosis, could produce effects that spread throughout a tissue.

## Introduction

Long range, intercellular calcium ($Ca^{2+}$) signaling is ubiquitous throughout the body. This mechanism of intercellular communication is utilized in nearly every stage of biological function, from the earliest moments of life, through metabolism, and even cell death, coordinating both single-cell systems and organ function alike. For example, $Ca^{2+}$ signaling helps activate an egg post fertilization, triggers insulin release in response to elevated blood sugar, and signals a cell to undergo apoptosis(1–4).

The conventional process of understanding long-range $Ca^{2+}$ signaling occurs as follows. First, an agonist binds to a cell and starts a process that eventually hydrolyzes phosphatidylinositol 4,5-bisphosphate ($PIP_2$) into inositol 1,4,5-trisphosphate ($IP_3$) and diacylglycerol (DAG)(5). $IP_3$ will then diffuse to the endoplasmic reticulum/sarcoplasmic reticulum (ER/SR), binding to $IP_3$ receptors ($IP_3Rs$) and release $Ca^{2+}$ from internal stores(6). $IP_3$ and $Ca^{2+}$ will then undergo ionic transport through gap junctions and propagate from one cell to the next, regenerating in each neighboring cell, allowing for these long-range signals to propagate to neighboring cells and allowing for coordinated dynamics in cellular systems(7–9).

This model is based heavily on Allbritton's seminal 1992 paper where they found that $IP_3$ diffuses much more freely and faster than $Ca^{2+}$. There, he introduced the idea of "range of action" for intercellular communication, the distance from a source over which ions and molecules can exert their influence(7). Based off the work done with xenopus models, $IP_3$ was initially understood to have a faster diffusion constant (~280 $\mu m^2/s$) and therefore further action range when compared to the diffusion rate of $Ca^{2+}$ (~3-65 $\mu m^2/s$). However, it is well known that $IP_3$ has a molecular weight of 412 g/mol(10), over ten times heavier than its $Ca^{2+}$ counterpart (40 g/mol) (11), and widely understood that heavier particles have slower diffusion rates than their lighter counterparts. Despite this, Allbritton concluded that the diffusion of $IP_3$, rather than the diffusion of $Ca^{2+}$ itself, was the mediator of long-distance intercellular signaling due to it having a further range of action. This led to the conclusion that $Ca^{2+}$ acts as a local signal of action, while $IP_3$ serves as a global signal between cells(8, 9, 12). Therefore, the molecular diffusion of $IP_3$ between gap junctions would serve as a regenerative mechanism to release $Ca^{2+}$ from internal stores, allowing the $Ca^{2+}$ signaling process to propagate across long distances.

However, recent work from Dickinson has shown that $IP_3$ has a much slower diffusion speed than previously thought (~10 $\mu m^2/s$), slower than the diffusion speed of $Ca^{2+}$ and measured rates of intercellular communication, therefore making it an insufficient mechanism for long-distance $Ca^{2+}$ propagation(11, 13, 14). Additionally, recent work has shown that completely disassembling gap junctions has no impact on $Ca^{2+}$ oscillation periods(15). Nevertheless, numerous experiments have shown that $IP_3$ is essential for calcium release and plays a key role in priming the release mechanism for the release of $Ca^{2+}$ (8). Together, these findings raise a fundamental question: if long-range calcium communication persists in the absence of rapid $IP_3$ diffusion and functional gap junctions, what alternative mechanism can account for the propagation of calcium signals across tissue?

In our study, we developed a computational model to demonstrate that local coupling between neighboring cells is sufficient to generate regenerative $Ca^{2+}$ oscillations that can propagate across a cell population. Using the Kuramoto model for nearest-neighbor coupling(16, 17), we treat each cell as an oscillator whose intrinsic frequency is set by its local $IP_3$ concentration ([$IP_3$]). These oscillations increase in frequency with applied [$IP_3$], allowing for synchronized calcium dynamics between cells. Specifically, this synchronization enables cells adhered to a physiologically stiff extracellular matrix (ECM) to amplify $Ca^{2+}$ oscillation frequency and contractility in neighboring cells on a soft ECM, propagating coordinated signaling across large distances. Our findings propose a new paradigm for calcium signaling and mechanotransduction in which local intercellular coupling, independent of molecular diffusion, serves as a regenerative mechanism for long-range calcium communication. This model can provide critical insights into how localized ECM remodeling can produce systemic effects, offering new insights into the early stages of disease progression associated with ECM alterations, such as asthma and fibrosis.

## Methods

All simulations were implemented and run in MATLAB R2020b (MathWorks, Natick, MA).

1. Receptor-level model of calcium signaling types

To reproduce the fundamental types of calcium signaling (Figure 1), we modeled a linear array of 10 $IP_3$ receptors along a single cell. Each receptor was treated as a dual-gated channel requiring occupancy of four $IP_3$ and four $Ca^{2+}$ binding sites to open, giving a nonlinear fourth-order ($p^4$) dependence of open probability on [$IP_3$] (Figure 1.B). Receptor opening was determined stochastically at each time step. A uniform random number was drawn, and the receptor opened and released $Ca^{2+}$ if the draw fell below the $p^4$-dependent open probability at local [$IP_3$] levels. Receptor opening and the associated $Ca^{2+}$ release were recorded at each position and time step to generate the space-time heatmaps in Figure 1. C–F, in which sparks, flashes, and waves emerge at low, intermediate, and high [$IP_3$], respectively. The open-probability curve in Figure 1.B was averaged over 10 independent stochastic simulations.

2. Oscillator model and $IP_3$-dependent refractory period

To capture calcium oscillation frequency (Figure 2), each receptor was represented as an oscillator with an absolute and a relative refractory period following each $Ca^{2+}$ release (Eqs. 1 and 2). The absolute refractory period was fixed at one time step, during which a receptor could not reopen. The relative refractory period was governed by an $IP_3$-dependent recovery rate $\omega$ (Eq. 1), with baseline recovery rate $\omega_0$= 0.2 and $IP_3$ sensitivity $\alpha = 5$. A receptor became available to reopen when its recovery phase $\theta$ reached $2\pi$ (Eq. 2). Frequency modulation (Figure 2. A–E) was simulated at [$IP_3$] = 1000, 2000, 3000, 4000, and 5000.

3. Kuramoto coupling model

To demonstrate how local coupling gives rise to synchrony (Figure 3), we simulated a system of N oscillators using the Kuramoto framework (Eqs. 3–7), with intrinsic frequencies drawn from a normal distribution and initial phases drawn uniformly from [0, $\pi$]. Coupling was defined by an adjacency matrix $\beta$ with entries in [0, 1], normalized by N, and the phase equations were integrated using MATLAB's ode45 adaptive solver. For the connectivity analysis (Figure 3. A–F), a specified fraction of all possible pairwise connections was activated at random. Results are shown for 1%, 10%, and 100% connectivity. For the nearest-neighbor configuration (Figure 3. G and H), oscillators were arranged in a ring and coupled only to their immediate neighbors.

4. Biological dual-stiffness model

To model long-range communication of substrate stiffness (Figures. 4 and 5), we simulated 15 confluent smooth muscle cells, each represented as an oscillator with the $IP_3$-dependent refractory dynamics described above, coupled only to its immediate neighbors in an open chain with no connection between the first and last cell. Local substrate stiffness set the local [$IP_3$] of each cell, which was held constant within each region throughout the simulation. Cell 1 was assigned to a stiff matrix ([$IP_3$] = 1600) and the remaining 14 cells to a soft matrix ([$IP_3$] = 1000). All oscillator phases were initialized to zero ($\theta = 0$). The nondimensionalized phase equation (Eq. 13) was integrated in dimensionless time $\tau$ using the forward Euler method, with dimensionless time step $d\tau = \omega_0 \, dt(\omega_0 = 0.2, \text{dt} = 0.01)$ for 10,000 steps per simulation. Heterogeneity was fixed at $\varepsilon = 25$, and the coupling strength K was set for each run to produce the target K/$\varepsilon$ ratio. Refractory-period measurements (Figure 4. A and B) were averaged over three independent simulations (N = 15 cells per simulation) and are reported as mean ± SD.

5. Nondimensionalization and characterization of collective behavior

The governing dynamics were nondimensionalized by rescaling time by the baseline recovery rate ($\tau = \omega_0 \, t$, Eqs. 9–13), yielding the dimensionless heterogeneity $\varepsilon = \alpha/\omega_0$ and coupling strength $K = C/\omega_0$, so that the collective behavior depends only on the ratio K/ε (Eqs. 11–13). To map the full range of behavior (Figure 5, Movies S1 and S2), we held ε = 25 fixed and varied K to sweep K/ε over 17 values ranging from 0.1 to 50. For each value of K/ε, the phase-locking value (PLV) and mean phase offset of each neighboring cell pair were computed as the magnitude and angle, respectively, of the time-averaged unit vector of their phase difference over the full simulation.

## Results

### 1. Fundamental calcium signaling dynamics emerge from the nonlinear gating properties of the $IP_3$ receptor.

The simplest forms of calcium signaling dynamics are sparks, flashes, and waves(18–20). These fundamental events represent the building blocks of intracellular and intercellular calcium communication. Each type represents a distinct scale of calcium release, from localized, short-lived events to large-scale propagating signals that span entire cells or tissues. Sparks are small, localized bursts of calcium release that occur within a confined region of a single cell. Flashes are when neighboring receptors open together, producing a coordinated $Ca^{2+}$ release that extends beyond a single localized site. They represent a larger and more organized signal than a spark but are still confined to a limited region of the cell. Waves represent the most organized form of calcium signaling, in which activity becomes synchronized and propagates across larger spatial domains, sometimes extending through entire tissues. Using only the current understanding of the $IP_3$ receptor and calcium release mechanisms, we can recreate these fundamental signaling dynamics. The $IP_3$ receptor functions as a dual-gated channel, containing four binding sites for both $IP_3$ and $Ca^{2+}$. In the absence of $IP_3$, the receptor's $Ca^{2+}$ binding sites favor an inhibitory configuration, preventing channel opening(21, 22). Once all four $IP_3$ sites are occupied, the receptor shifts from an inhibitory to a stimulatory configuration, increasing the affinity for $Ca^{2+}$ binding. Only when all four $Ca^{2+}$ sites are subsequently filled does the channel open, releasing calcium from the internal stores of the cell(23–25). This allows for a dual-gated process, wherein $IP_3$ is necessary, and even primes, for $Ca^{2+}$ release that leads to calcium-induced calcium release (CICR) (Figure 1.A)(26, 27). This mechanism gives rise to a nonlinear, fourth order ($p^4$) dynamic that governs $IP_3$ -mediated CICR (Figure 1.B). At low concentrations of $IP_3$, the probability of receptor opening is nearly zero. Once [$IP_3$] surpasses a critical threshold, the probability rises sharply, leading to a rapid increase in the fraction of open receptors. Utilizing this dynamic alone, at a fixed [$Ca^{2+}$] and varying [$IP_3$], the three signaling behaviors (sparks, flashes and waves) of $Ca^{2+}$ signaling occur at low, medium and high concentrations of [$IP_3$] respectively (Figure 1.C-F).

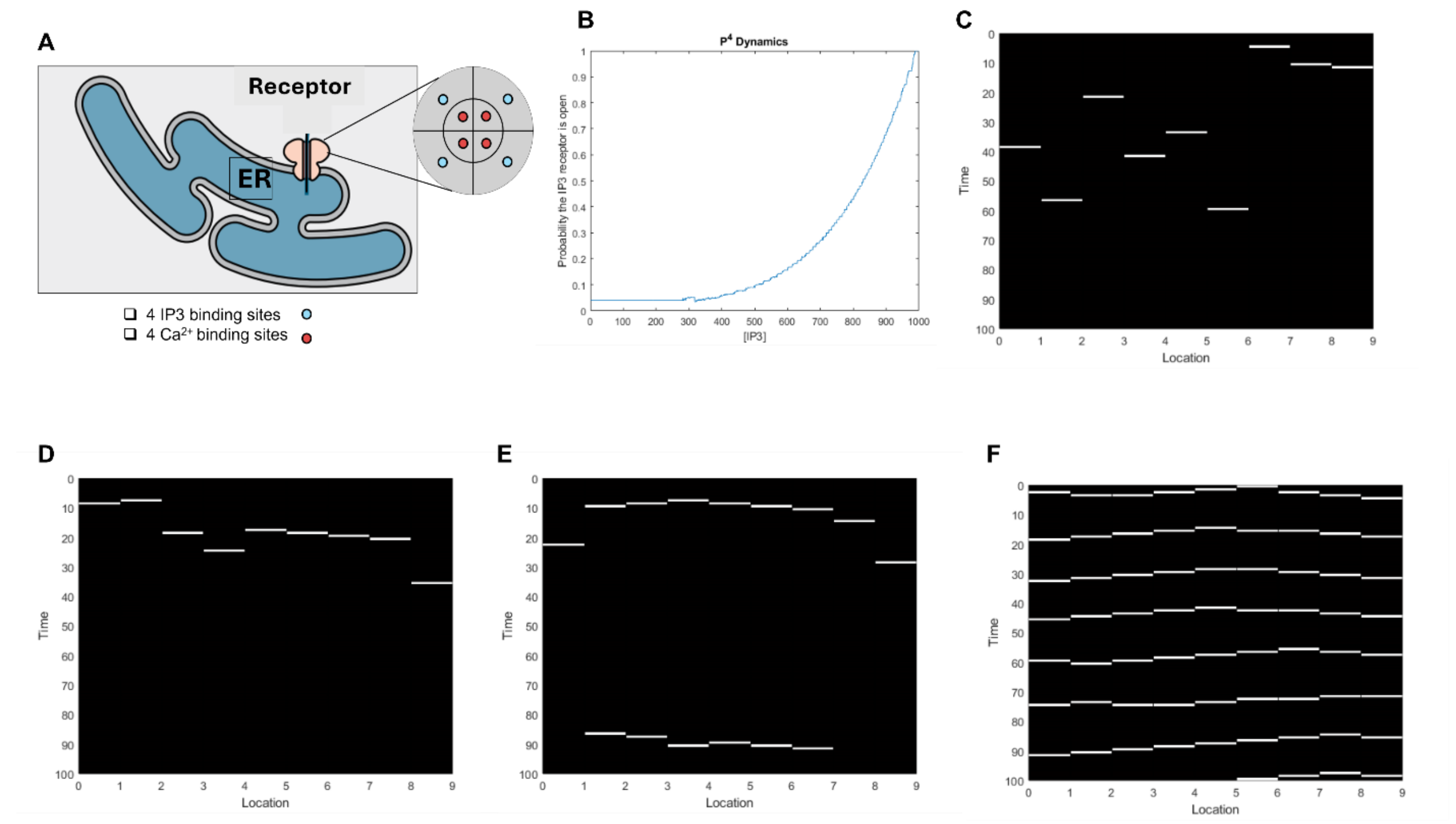

<u>Figure 1: The nonlinear gating of the $IP_3$ receptor gives rise to the fundamental types of calcium signaling</u>

A) Schematic of the ER-based $IP_3$ receptor and its gating mechanism. The receptor is a dual-gated channel with four subunit binding sites for $IP_3$ (Blue) and four for $Ca^{2+}$ (Red). The inset shows a simplistic view of the binding site architecture. Only when all four $IP_3$ binding sits *AND* all four $Ca^{2+}$ binding sites are bound can the channel open, releasing $Ca^{2+}$ from the internal stores of the cell. This dual-gated mechanism enables calcium-induced calcium release, in which $IP_3$ binding primes the receptor and $Ca^{2+}$ binding triggers the release.

B) Normalized fraction of $IP_3$ receptors in an open state as a function of $IP_3$ concentration averaged over 10 simulations. The relationship follows a nonlinear, fourth order ($p^4$) dynamic. At low [$IP_3$], the probability of the receptor opening is near zero. However, as [$IP_3$] increases, the probability of the receptor opening increases, enabling $Ca^{2+}$ release. This gating dynamic alone gives rise to the three fundamental types of calcium signaling: sparks at low [$IP_3$], flashes at intermediate [$IP_3$], and waves at high [$IP_3$].

C-F) Heatmaps showing calcium release patterns arising from activation of receptors arranged linearly along a cell as a function of [$IP_3$]. The x-axis denotes receptor position along the cell and the y-axis denotes time, where each step corresponds to one iteration of the simulation. White regions indicate receptor opening and the associated $Ca^{2+}$ release.

C) At low levels of [$IP_3$] only a few isolated receptors open, producing localized sparks.

D) As [$IP_3$] increases receptors start to activate in a partially coordinated manner which gives rise to flashes, which are coordinated $Ca^{2+}$ release over a broader region of the cell.

E) Further increases in [$IP_3$] concentration produce a combination of flashes and short waves.

F) At high [$IP_3$], receptor activity becomes fully coordinated and gives rise to waves, which are $Ca^{2+}$ signals that travel along the length of the cell.

## 2. Calcium oscillation frequency is governed by an $IP_3$-dependent refractory period that enables frequency-encoded signaling.

Previous studies have shown that calcium signaling is far more complex than simple molecular diffusion. For example, in airway smooth muscle, higher concentrations of an agonist produce more frequent calcium oscillations(28). This behavior suggests that calcium functions as a carrier wave that encodes and transmits information between cells. To capture this relationship, we describe the underlying dynamics using a mathematical model that links receptor behavior and oscillation frequency. A key factor in these dynamics is the refractory period, the brief recovery phase that follows each calcium release. During this time, the $IP_3$ receptors and associated channels are temporarily inactive as calcium is reabsorbed into internal stores and the system resets. The length of this refractory period determines how quickly another release can occur. In this model, we propose that each receptor experiences both an absolute and a relative refractory period. The absolute refractory period prevents a receptor from reopening immediately after $Ca^{2+}$ release, during which the receptor cannot reopen under any conditions. The relative refractory period can be defined by:

$$\omega = \omega_0 + \alpha\,[IP_3]\ Eq.\ 1$$

where $\omega$ is defined as the $IP_3$ dependent frequency of receptor recovery, $\omega_0$ is a constant baseline recovery rate, and $\alpha$ is a constant relating to how strongly $IP_3$ concentration influences the refractory period. The phase of the system would be defined as:

$$\theta_{(x,t)} = \theta_{(x,t-1)} + \omega\ dt\ Eq.\ 2$$

where $\theta$ is defined as the phase of the oscillator recovery cycle. Only when θ reaches 2π, the receptor becomes available to open again, completing the relative refractory period.
Together, these equations allow the population of receptors to vary their calcium release frequency based solely on the local $IP_3$ concentration. As $IP_3$ increases, $\omega$ increases, shortening the relative refractory period and producing higher-frequency calcium oscillations (Figure 2.A-C).

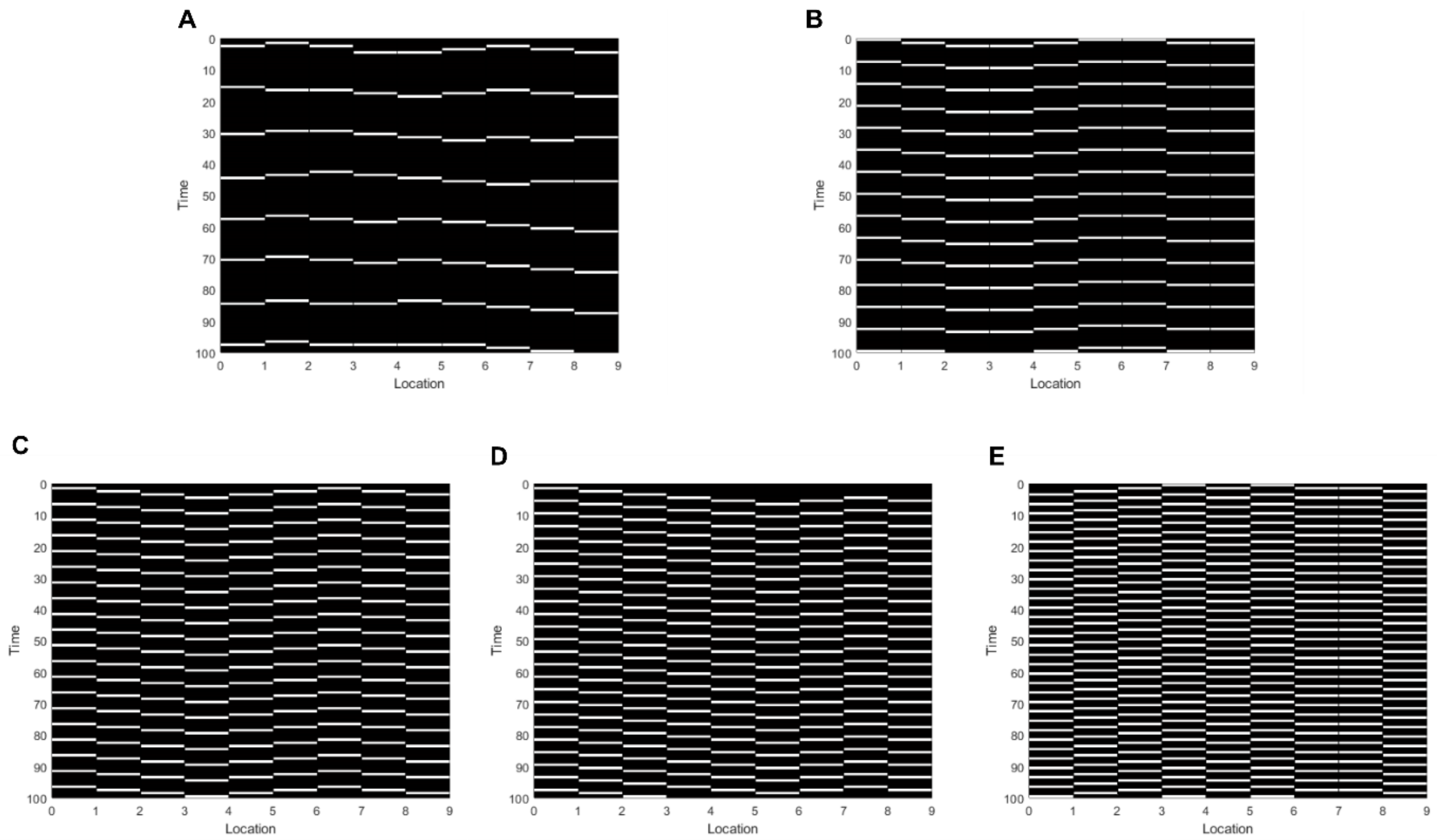

Figure 2: $IP_3$-dependent refractory period modulates the frequency of $Ca^{2+}$ oscillations

Frequency modulation of $Ca^{2+}$ oscillations by $IP_3$ concentration.

Incorporating both an absolute and relative refractory period into the model enables oscillation frequency to be modulated as a function of $[IP_3]$. Panels A–E show spatiotemporal heatmaps of $Ca^{2+}$ release from receptors distributed linearly along a cell at $[IP_3]$ = 1000, 2000, 3000, 4000, and 5000, respectively. Receptor position is shown on the x-axis and time on the y-axis, with white regions indicating receptor opening and associated $Ca^{2+}$ release. As $[IP_3]$ increases from panel to panel, the frequency of $Ca^{2+}$ oscillations rises accordingly.

## 3. Synchrony and Communication

These refractory-period dynamics establish the timing of individual receptor activation and the frequency of collective oscillations. However, long-range calcium-based communication depends on the idea that different cells influence one another through direct interactions. In previous work from our group we demonstrated that smooth muscle cells (SMCs) communicate the underlying stiffness of the extracellular matrix (ECM), where SMCs on a stiff matrix (13 kPa) can influence the frequency of oscillations of cells on a soft matrix (0.3 kPa) over 8 cell lengths away(15). Additionally, SMCs on a soft matrix closer to the SMCs on a stiff matrix had a higher oscillation frequency than those further away. This suggests that coordination arises from interactions between adjacent cells.

To this end, we developed a model in which each cell is treated as an individual oscillator. This representation arises from the periodic, repeating nature of the calcium release cycle, which follows a repeating pattern of activation, refractory recovery, and reactivation governed by the $IP_3$-dependent dynamics described above(29). The frequency of this cycle is determined by the local $IP_3$ concentration, making each cell's oscillatory behavior a direct reflection of its mechanical environment, as stiffer substrates leads to faster rates of $Ca^{2+}$ signaling(15). However, the relationships described thus far only describe how a cell responds to its own environment and does not account for how neighboring cells influence one another. To address this, we incorporate a coupling mechanism into the model that enables cells to "communicate" and synchronize to support the propagation of calcium signals throughout the tissue. Treating each cell as an oscillator allows us to reframe this question as a coupled oscillator problem, where the key question is how the frequency, and by extension phase, of one cell's calcium cycle influences that of its neighbors.

We therefore incorporate the Kuramoto model wherein local oscillators can couple and synchronize across space(16). Using the simple case of 2 oscillators, we can describe the system where:

$$y_1 = \sin(\omega t + \theta_1) \; Eq.\ 3$$
$$y_2 = \sin(\omega t + \theta_2) \; Eq.\ 4$$

where $y_1$ and $y_2$ are two oscillators which have the same initial frequency but a different initial phase. The evolution of the phase of these oscillators can be described by

$$\frac{d\theta_1}{dt} = \omega + \beta \sin(\theta_2 - \theta_1) \; Eq.\ 5$$
$$\frac{d\theta_2}{dt} = \omega + \beta \sin(\theta_1 - \theta_2) \; Eq.\ 6$$

where $\beta$ is an adjacency matrix whose entries lie in the interval [0, 1] that represents which oscillators in the system are coupled. When $\beta$ is zero, the two oscillators act independently. However, when the connected fraction $\beta > 0$, the second term in Eqs. 5 and 6 begin to adjust the phase dynamics so that the leading oscillator slows down and the lagging oscillator speeds up. As their phase difference decreases, their instantaneous frequencies must also shift accordingly, causing the oscillators to converge toward a common frequency. When $\beta = 1$, the oscillators are perfectly connected. Extrapolating from this, the Kuramoto model can be extended to a set of oscillators or cells where,

$$\frac{d\theta_i}{dt} = \omega_{i,t-1} + C * \sum_{j=1}^{N} \beta_{i,j} \sin(\theta_j - \theta_i) \; Eq. 7$$

Here β remains the adjacency matrix delineating which cells are connected, with entries in [0, 1], while C is a coupling coefficient setting the overall strength of interaction. Separating the two allows the topology of connections and the strength of coupling to be varied independently. Using this model, we simulated a system of 15 confluent smooth muscle cells (SMCs), each represented as an individual oscillator. Pairwise interactions between oscillators $i$ and $j$ were established by setting $\beta_{i,j}$ equal to 1. To assess how connectivity influences collective behavior, we randomly activated a specified percentage of these possible connections across a series of simulations.

As expected for systems of coupled oscillators, increasing the fraction of active connections drove the system toward synchronization(30). At low connectivity, synchronization was minimal. In a network in which only 1% of all possible pairs interact ($\beta_{2,15} = 1$), the oscillators display little coordinated activity (Figure 3 A-B). As the fraction of active connections increased, the system progressively moved toward synchronization. Comparing networks at 10% and 50% connectivity (Figure 3 C-D), the oscillatory dynamics became markedly more coherent (Figure 3 E-F), showing a direct relationship between connection density and the degree of global synchrony.

We also examined a nearest-neighbor configuration, in which oscillators were arranged in a circle and coupled only to their immediate neighbors (Figure 3 G). This topology produced a phase-locked traveling wave in which neighboring cells oscillate in phase but with a consistent temporal delay, a pattern we refer to as "offset synchrony" (Figure 3 H). This behavior closely resembles calcium wave propagation observed in SMC experiments, where one cell releases calcium and, in turn, signals its neighbor to do the same, initiating a sequential cascade of communication along the cell population(15).

To model long distance intercellular communication, we applied a nearest-neighbor configuration (Figure 3 G) by incorporating the Kuramoto model into our existing framework, where each cell influences only its immediate neighbors according to Eq. 7. Experimental evidence has shown that cells on stiffer substrates exhibit higher frequencies of $Ca^{2+}$ oscillations, and that $IP_3$ is a well-established driver of $Ca^{2+}$ release(5, 8, 15, 28, 31). We therefore introduce a dual-stiffness system in which local ECM stiffness determines the steady-state $IP_3$ concentration assigned to each cell. Cells positioned on the stiff region are assigned a higher [$IP_3$], and therefore a higher intrinsic frequency (ω) than cells in the soft region. Both regions receive a constant [$IP_3$], as intercellular $Ca^{2+}$ release has been shown to occur even when [$IP_3$] is held constant(32). Through coupling, the higher-frequency cells on the stiff region pull their immediate soft-region neighbors into step by advancing their phases, causing the cells on the soft matrix to increase their oscillation frequency, consistent with experimental observations(15).

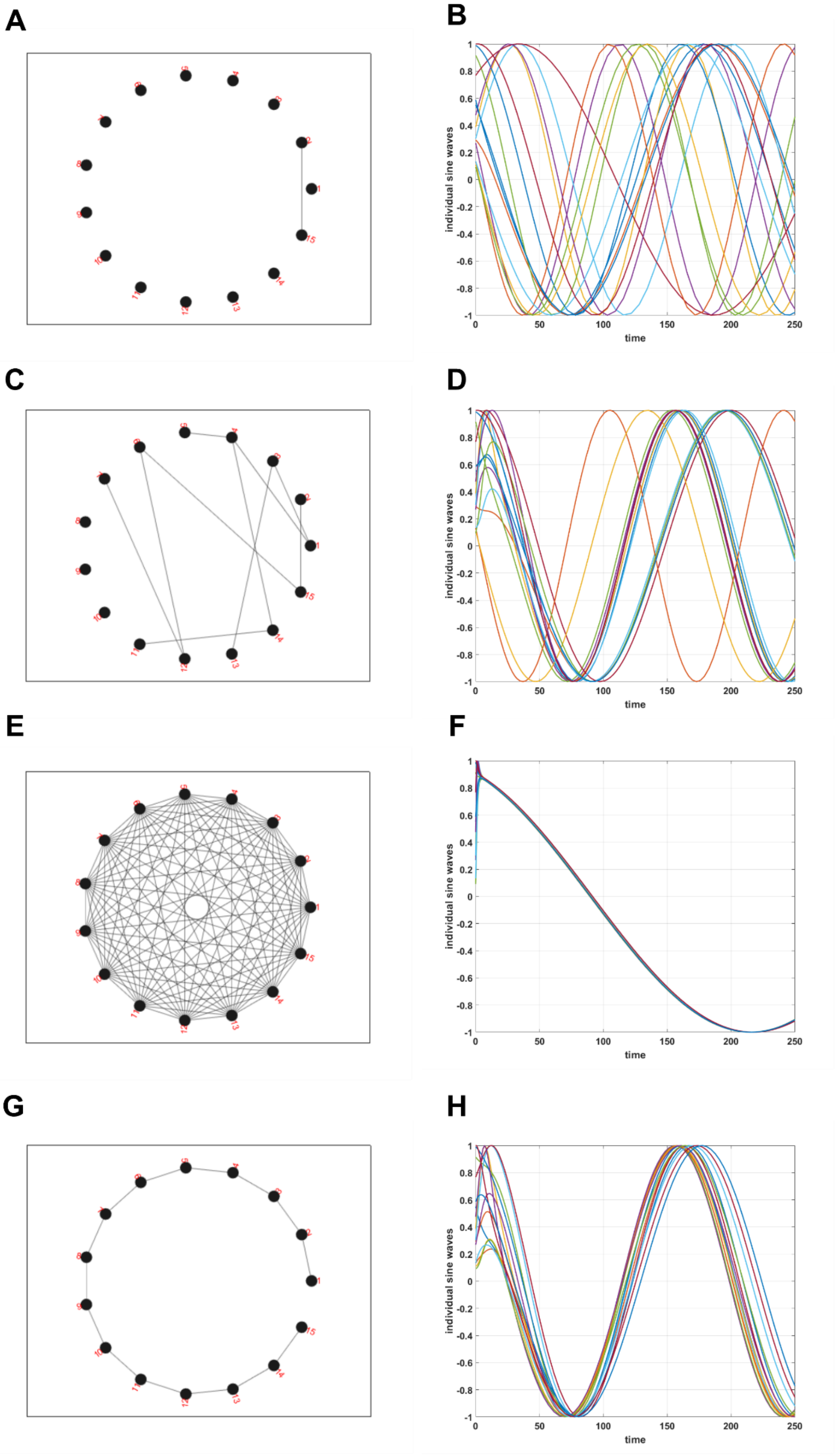
A
B
C
D
E
F
G
H
individual sine waves
time
1
0.8
0.6
0.4
0.2
0
-0.2
-0.4
-0.6
-0.8
-1
0
50
100
150
200
250

Figure 3: Kuramoto phase coupling drives the emergence of synchrony and intercellular communication

Utilizing the Kuramoto model, we simulated the effects of increasing the number of connected pairs on the emergence of synchrony and intercellular communication.

A) Network diagram of a 1% connected system in which only a single pair of oscillators is coupled ($\beta_{2,15} = 1$). The black dots represent smooth muscle cells which are modeled as oscillators. A line between two dots represent that the two oscillators are connected and interacting.
B) Phase diagram showing minimal synchronization under 1% connectivity, with oscillators behaving largely independently. Each color represents a different oscillator.
C) Network diagram of a 10% connected system.
D) Phase diagram showing the emergence of partial synchronization at 10% connectivity. Each color represents a different oscillator.

E) Network diagram of a 100% connected system.
F) Phase diagram showing complete global synchronization under full connectivity. As the number of connections between cells increases, the more the system synchronizes their oscillations.
G) Nearest-neighbor configuration in which cells are connected linearly. Oscillator interactions are restricted such that each cell can only influence the phase and resulting frequency of its immediate neighbors, similar to a line of cells.
H) Phase diagram of the nearest-neighbor configuration showing offset synchrony, in which each cell oscillates slightly out of phase with its neighbor, producing a sequential cascade of calcium release. Each color represents a different oscillator.

At the start of the simulation, the cell on a stiff matrix (Figure 4.A, Left, Red) oscillates at a faster rate than the cells on the soft matrix (Figure 4.A, Right, Blue). As the simulation progresses, the cells on the soft matrix begin to exhibit offset synchrony, indicating the emergence of intercellular communication (Figure 4.B). When the stiff matrix cell releases calcium, its nearest neighbor responds with its own calcium release, initiating a sequential cascade along the line of cells. Three distinct spatial regimes emerge as distance from the stiff cell increases. Cells closest to the stiff-matrix cell exhibit full offset synchrony, firing sequentially in direct response to the stiff cell. At intermediate distances, cells display partial offset synchrony, in which small groups of neighbors synchronize to one another while still participating in the broader communication cascade. At approximately eight cell lengths from the stiff-matrix cell, the cells fully synchronize with their immediate neighbors and no longer exhibit a phase relationship with the stiff cell, as the coupling is too weak at this distance to influence their dynamics (Figure 4.B-C). This result is consistent with experimental data reported by Stasiak et al(15). In their study, cells on a soft matrix maintained oscillation periods statistically similar to those on the stiff matrix up to 800 µm away. Beyond this distance, oscillation periods diverged significantly.

Notably, our model produces a finite spatial range of influence consistent with experimental observations(15). Cells approximately eight cell lengths (~800 µm)(33, 34) from the stiff-matrix cell still show evidence of communication with the stiff matrix cell, suggesting that nearest-neighbor coupling alone is sufficient to explain the observed range of intercellular communication.

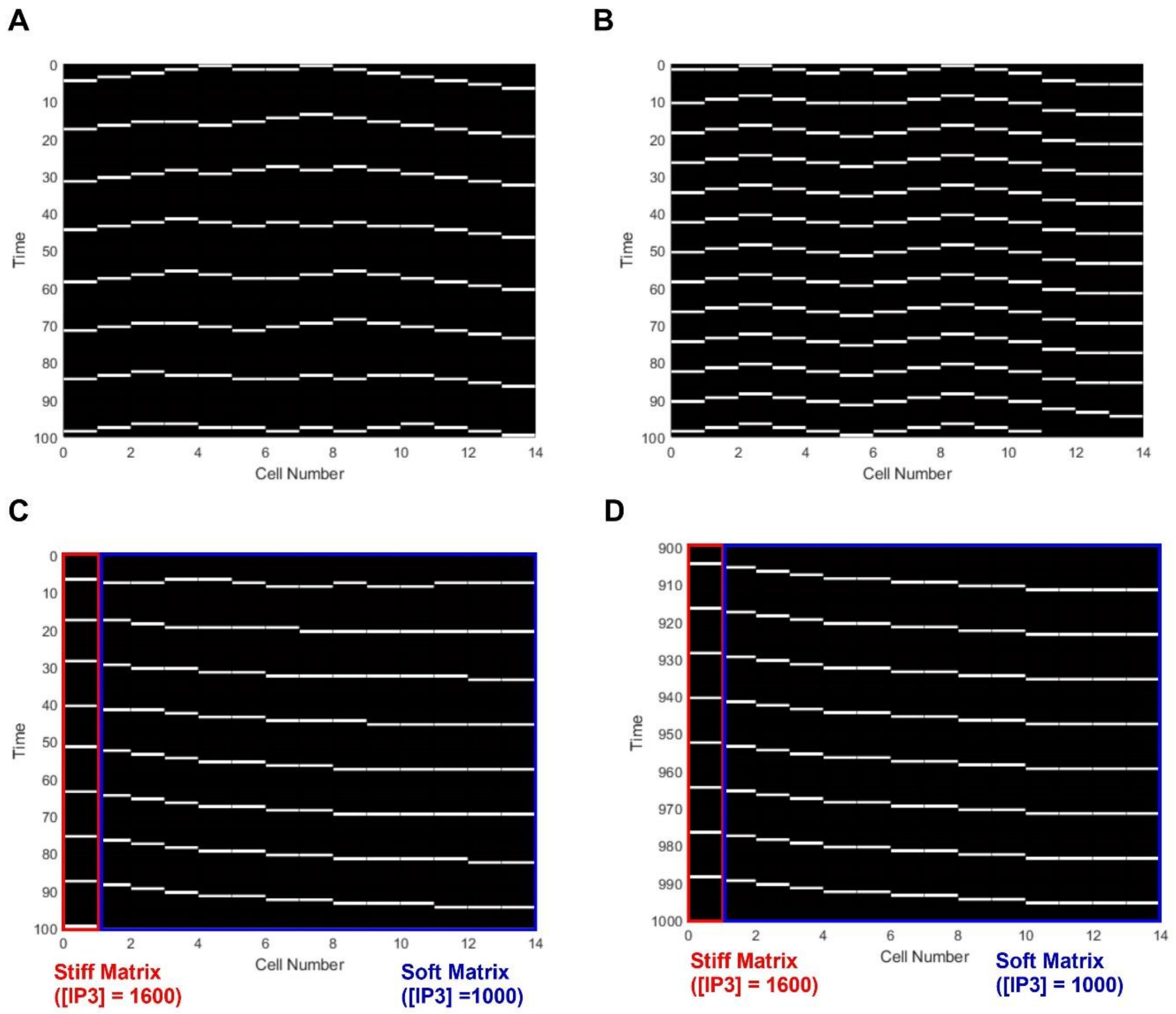

Figure 4: $IP_3$-dependent oscillation frequency and nearest-neighbor coupling produce an offset traveling wave with a finite spatial range

A) Refractory period of a line of uncoupled cells at $[IP_3] = 1000$. Cells exhibit a mean refractory period of 12.51±0.23 steps between $Ca^{2+}$ release (N = 15 cells, averaged over 3 separate simulations).
B) The same system at $[IP_3] = 1600$. Here, cells exhibit a mean refractory period of 7.01±0.03 steps between $Ca^{2+}$ release (mean±SD, N = 15 cells, averaged over 3 separate simulations). This demonstrates the $IP_3$-dependent modulation of oscillation frequency.
C) Spatiotemporal heatmap of a dual-stiffness, coupled nearest-neighbor model at the start of the simulation. Fifteen cells are arranged in a line coupled only to their immediate neighbors, with no connection between the first and last cell. Cell 1 (red) is positioned on a stiff matrix and therefore assigned a higher $[IP_3]$ ($[IP_3] = 1600$), while the remaining 14 cells (blue) are on a soft matrix with a lower $[IP_3]$ ($[IP_3] = 1000$). Early in the simulation, the cells operate independently, with the stiff-matrix cell oscillating at a higher frequency and the soft-matrix cells oscillating more slowly.
D) The same system at steady state. As intercellular communication drives the system toward synchronization, three distinct spatial regimes emerge. Cells nearest to the stiff-matrix cell exhibit full offset synchrony, firing sequentially in a wave-like cascade in direct response to the stiff cell. At intermediate distances, cells display partial offset synchrony, in which small groups of neighbors synchronize to one another while still participating in the broader communication cascade. At approximately eight cell lengths from the stiff-matrix cell, cells enter a regime of no communication as the coupling is too weak, allowing the cells to fully synchronize with their immediate neighbors but no longer exhibiting a phase relationship with the stiff cell.

4. **Collective behavior is governed by the ratio of coupling to heterogeneity**

Our model in its current form contains three independent parameters: the baseline recovery rate $\omega_0$, the $IP_3$ sensitivity α (from Eq. 1), and the coupling strength C (from Eq. 7). Substituting the $IP_3$-dependent recovery frequency of Eq. 1 into Eq. 7 gives

$$\frac{d\theta_i}{dt} = (\omega_0 + \alpha * [IP_3]) + C * \sum_{j=1}^{N} \beta_{i,j} \sin(\theta_j - \theta_i) \; Eq. 8$$

The collective behavior of the system emerges from the interplay of these three parameters. To identify which of their combinations actually govern the dynamics of the system, we nondimensionalize the system and measure time in the natural units of the oscillation itself. Therefore, we measure time relative to the baseline recovery rate $\omega_0$, the one rate common to every cell, by defining a dimensionless time

$$\tau = \omega_0 * t \; (Eq. 9)$$

so that one unit of τ corresponds to one baseline recovery cycle. Substituting this into Eq. 7 and dividing through by $\omega_0$ gives

$$\frac{d\theta_i}{d\tau} = 1 + \frac{\alpha}{\omega_0} * [IP_3] + \frac{C}{\omega_0} * \sum_{j=1}^{N} \beta_{i,j} \sin(\theta_j - \theta_i) \; Eq. 10$$

Two parameter groupings emerge from this, $\alpha/\omega_0$ and $C/\omega_0$, which we define as the dimensionless heterogeneity (ε) and coupling strength (K) where:

$$\varepsilon = \frac{\alpha}{\omega_0} \; Eq. 11$$

$$K = \frac{C}{\omega_0} Eq. 12$$

ε sets how strongly a cell's oscillation frequency responds to its $IP_3$ level, measured relative to the baseline recovery rate. Because $[IP_3]$ varies from cell to cell, driven by differences in the stiffness of the surrounding matrix, ε determines how different the cells are in their natural frequencies. A large ε turns even small differences in $IP_3$ into large differences in frequency, so the population becomes highly heterogeneous. A small ε does the opposite, leaving the cells with nearly the same natural frequency and recovery rate. K sets how strongly a connected neighbor pulls on a cell's oscillation, again relative to the baseline rate. A large K means the coupling is strong compared to a cell's own pace, so the cell is readily drawn toward the phase of its neighbors. A small K means the coupling is weak, so each cell follows its own rhythm and is barely affected by those around it. In short, K controls how tightly the cells are bound to one another.

With these definitions, Eq. 7 takes its final dimensionless form

$$\frac{d\theta_i}{d\tau} = 1 + \varepsilon * [IP_3] + \mathrm{K} * \sum_{j=1}^{N} \beta_{i,j} \sin(\theta_j - \theta_i) \; Eq. 13$$

Although ε and K are independent parameters, the collective behavior of the system does not depend on either one alone. Rather, the two parameters compete with one another. The heterogeneity term $\varepsilon * [IP_3]$drives cells apart, pushing each toward its own intrinsic frequency, while the coupling term $\mathrm{K} * \sum_{j=1}^{N} \beta_{i,j} \sin(\theta_j - \theta_i)$ draws them together, pulling each cell toward the phase of its neighbors. The outcome is therefore set by the balance between these competing influences by the ratio K/ε rather than by K or ε individually. Equivalently, scaling ε and K together leaves the dynamics unchanged, so only their ratio governs the behavior. This governing ratio emerges directly from the biological structure of our model. Since both the heterogeneity ($\varepsilon = \alpha/\omega_0$)and the coupling ($\mathrm{K} = C/\omega_0$) are scaled by the same baseline recovery rate, the dynamics reduce to the single quantity K/ε = C/$\alpha$, the ratio of coupling strength to $IP_3$ sensitivity. The general principle that such a ratio governs entrainment in pacemaker-driven oscillator networks has been established previously, but here it arises naturally from a model grounded in receptor biology(35).

To map the full range of behaviors the system can produce, we systematically varied K/ε and characterized the resulting dynamics (Movie S1). Rather than a simple progression from disorder to order, the system passes through a coherent wave-propagating regime between disorder and full synchrony.

At low K/ε, the cells remain coupled, but the coupling is too weak relative to the large frequency difference between the stiff-matrix cell and the soft-matrix cells to entrain them. The soft cells, which share similar intrinsic frequencies, synchronize closely among themselves into a coherent block, while the faster stiff cell oscillates largely on its own. At the lowest values of K/ε the coupling has no discernible effect. Here, the soft block fires uniformly, showing no influence from the stiff cell. As K/ε rises, the stiff cell begins to perturb the soft cells nearest the boundary, but the frequency mismatch remains too large for them to lock. Rather than holding a fixed phase relationship, these cells periodically fall a full cycle behind the faster stiff cell. In this sub-threshold regime, the coupling is present but too weak to entrain the population, and information from the stiff cell propagates poorly or not at all (Figure 5.A).

As K/ε increases into an intermediate range, coupling becomes strong enough for the stiff cell to entrain its neighbors while still permitting a consistent phase difference between them. This produces the offset traveling wave described above (Figure 4.C-D). Calcium release propagates outward from the stiff cell in a fixed sequence, with the offset largest near the stiff cell and relaxing to local synchrony farther along the line. This wave-propagating regime is where communication succeeds, reproducing the finite spatial range of influence observed experimentally. Within this regime, the spatial extent of the wave depends on K/ε. At the upper end, the offset persists across much of the line, extending the apparent range of influence beyond the experimentally observed distance (Figure 5.B, Movie S1). At sufficiently high K/ε, however, the coupling overwhelms the frequency heterogeneity entirely. The cells are pulled toward a common phase, the spatial offset collapses, and the population transitions from a propagating wave into near-simultaneous, in-phase synchrony (Figure 5.C).

The phase-locking structure underlying these regimes becomes clearer when we represent each neighboring pair of cells by its phase-locking value (PLV) and mean offset (Movie S2). This view separates two properties that the space-time maps blend together. The first is whether a pair is locked, meaning it holds a stable phase relationship over time. The second is the size of the offset between the pair if they are locked. At low K/ε, the stiff-soft substrate pair sit near the center showing little stable locking, while the soft substrate pairs are synchronized at the boundary. This fits the weakly coupled regime, where the soft cells follow one another but not the stiff driver. As K/ε increases, the pairs move outward and become phase-locked, but they settle at nonzero offsets rather than at zero. The cells lock together while staying phase-shifted, which is the defining feature of the offset wave. This tells us that the traveling wave reflects genuine phase-locking between cells and not just cells that happen to fire one after another. At higher K/ε, the locked pairs rotate toward zero offset, marking the shift toward full in-phase synchrony (Figure 5.C).

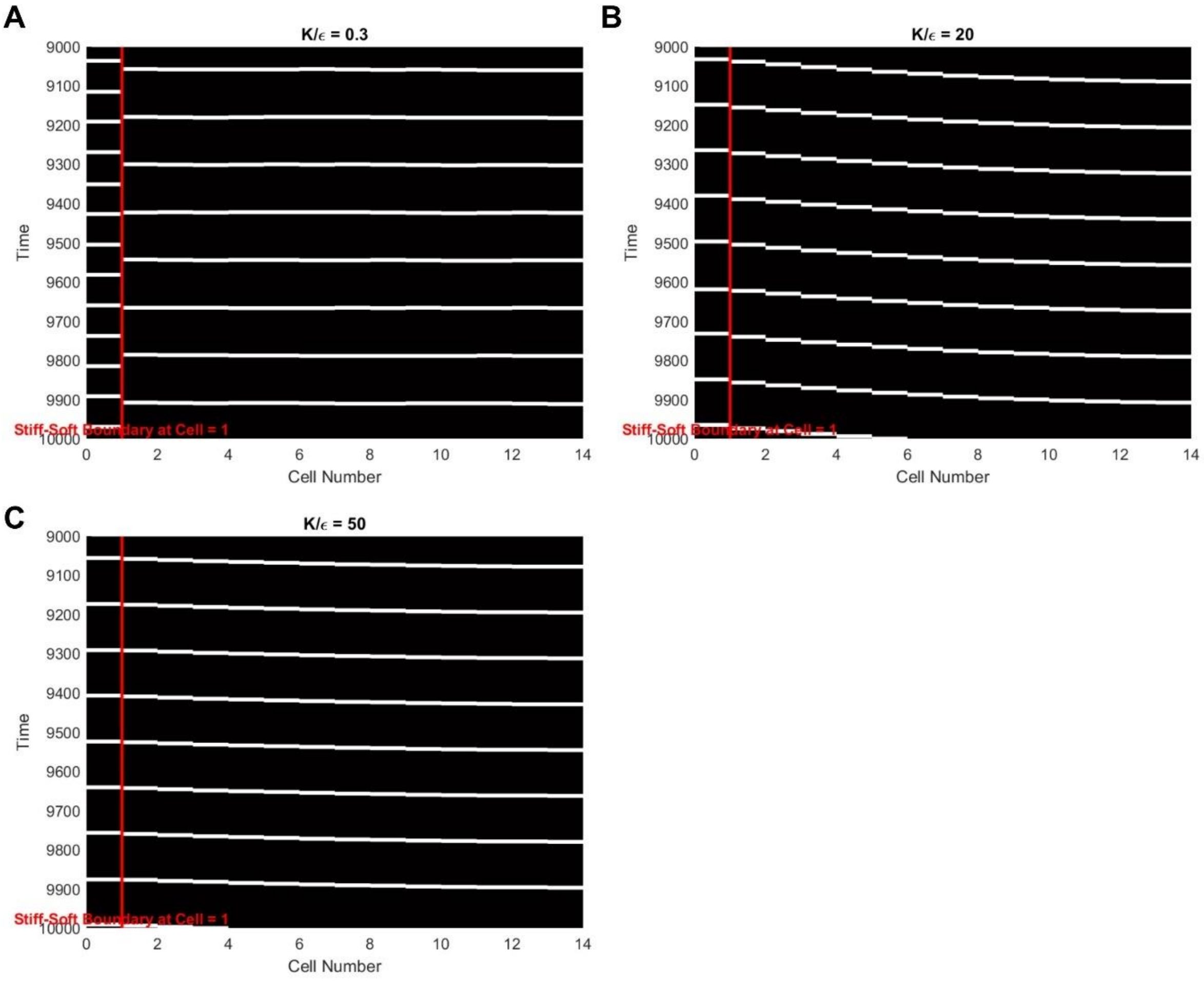

<u>Figure 5: The ratio of coupling to heterogeneity governs the transition from incoherence to wave propagation to full synchrony</u>

A) At low K/ε (weakly coupled regime), the soft cells fire as a synchronized block while the faster stiff cell oscillates independently and substrate stiffness information does not propagate.
B) At intermediate K/ε (wave-propagating regime), a coherent offset wave propagates outward from the stiff cell, with the phase offset largest near the stiff cell and relaxing to local synchrony farther along the line.
C) At high K/ε (strong-coupling regime), coupling dominates the frequency heterogeneity and pulls the cells toward a common phase. The spatial offset that defined the traveling wave largely disappears, and the population approaches full in-phase synchrony, firing nearly simultaneously across the line.

**Movie S1:**
Space-time map of calcium release across the line of cells as K/ε increases (Movie S1). Each row shows the firing pattern of the cell population over time, with the stiff-matrix cell at the left (red boundary line) and soft-matrix cells extending rightward. As K/ε increases, the population transitions from a weakly coupled state, in which the soft cells synchronize among themselves independently of the stiff cell, to a wave-propagating state, in which calcium release propagates outward from the stiff cell as a coherent offset wave.[1]

**Movie S2:**
Phase-locking representation of the same K/ε sweep (Movie S2). Each point represents a neighboring pair of cells, positioned in the complex plane by two quantities, 1) its distance from the center gives the phase-locking value (PLV), a measure of how consistently the pair maintains a fixed phase relationship over time, and 2) its angle gives the mean phase offset between the pair. Points near the center are unlocked (no stable phase relationship). Points near the rim are phase locked, with those at zero angle firing in phase and those at nonzero angle maintaining a fixed offset. Points are colored by position along the cell line. As K/ε increases, the points migrate from the center outward to the rim and rotate toward zero angle, tracing the progression from incoherence (unlocked) through the offset wave (locked but phase-shifted) to full synchrony (locked and in phase).

**Discussion**

Calcium signaling is one of the most tightly regulated communication systems in biology, governing processes ranging from egg fertilization and metabolism to contraction and apoptosis. Yet despite decades of study, classical models that rely on rapid $IP_3$ diffusion cannot fully explain the long-range propagation of calcium signals or the persistence of oscillations after gap junction inhibition. These limitations suggest that an alternative or complementary mechanism may underlie intercellular calcium communication.

In this study, we show that an $IP_3$-dependent oscillator model combined with a Kuramoto coupling framework, is sufficient to generate and propagate calcium oscillations across a cell population without requiring fast molecular diffusion. The model relies on two principles. The first is that the refractory period governing each oscillator is relative and is determined by local $IP_3$ concentration. This means that cells on a stiffer substrate, which maintain higher [$IP_3$], recover from refractory periods more quickly and therefore oscillate at higher intrinsic frequencies. Unlike models that treat refractory dynamics as a fixed property, this $IP_3$-dependent recovery rate allows oscillation frequency to emerge naturally from the local mechanical environment. The second is that coupling between nearest neighbors draws soft-region cells forward in phase, increasing their instantaneous frequency until offset phase locking emerges. Together, these dynamics offer a simple mechanistic explanation for how calcium signals propagate across tissue. Importantly, the model reproduces the spatial boundary of communication observed experimentally by Stasiak et al. which suggests that nearest-neighbor coupling alone is sufficient to explain long-range substrate stiffness influence(15).

[1] FigShare link to S1 and S2: https://figshare.com/s/aaa80049f63d78541dc4

A key consequence of this framework is that calcium propagation is inherently regenerative. Rather than relying on a diffusing signal that decays with distance, each cell independently generates its own calcium release event in response to phase coupling with its neighbor. The signal is actively renewed at every step which allows for robust long-range communication through purely local interactions. This behavior is fundamentally different from diffusion-based models and suggests that coupling alone can sustain calcium propagation across large cell populations.

Converting the model to a dimensionless form shows that its collective behavior is governed not by the coupling strength nor the frequency heterogeneity on their own, but by the balance between them, which we capture in the single ratio $K/\varepsilon$. This simplification is useful because it means the same behavior can arise from many different combinations of underlying parameters, so long as their ratio is preserved. As we varied $K/\varepsilon$, the system moved through a clear progression. When coupling was weak relative to the frequency differences between cells, the soft cells synchronized among themselves but did not track the faster stiff cell, and substrate stiffness information failed to propagate. As the ratio increased, the stiff cell became able to entrain its neighbors while still holding a consistent phase difference between them, producing the offset traveling wave that underlies long-range communication in our model. At higher ratios still, the coupling was strong enough to pull the cells not just to a common frequency but toward a common phase, so that the population approached full synchrony and the spatial offset largely disappeared. This progression suggests that successful communication occupies an intermediate regime, where coupling is strong enough to link cells together but not so strong that it collapses their distinct phases into uniform, simultaneous firing. As the traveling wave appears across a range of $K/\varepsilon$ rather than at a single value, wave propagation itself is a robust feature of the coupling and heterogeneity balance rather than a finely tuned outcome, although the spatial extent of the wave does depend on where within this range the system sits.

These findings carry significant implications for understanding disease. Localized increases in ECM stiffness are a hallmark of conditions such as asthma and fibrosis. Our model raises the possibility that such mechanical changes could spread altered signaling and connection into adjacent healthy tissue through frequency entrainment, whereby a stiff region drives neighboring cells towards higher oscillation frequencies. This offers a possible framework for understanding how localized ECM remodeling might produce systemic effects and contributes to the spatial patterning of tissue dysfunction during early stages of disease progression.

This work opens several avenues for future investigation. In the current model, substrate stiffness is represented through its effect on $[IP_3]$ and $[IP_3]$ is treated as a static parameter set mathematically. To bridge the gap between its mathematical framework and the underlying biology, incorporating dynamic $IP_3$ regulation calibrated to experimentally measured concentrations and substrate stiffnesses would allow the model to capture the temporal signaling behavior observed in living cells. Additionally, while our simulations demonstrate the sufficiency of coupling as a signaling mechanism, the model still treats intercellular connectivity as an abstract coupling term ($\beta$). Identifying the physical basis of this coupling, whether through direct mechanical force transmission, cadherins, integrins, or other pathways, is an important next step. Berridge's work on force-induced $IP_3$ production offers a promising starting point, and targeted experiments could determine how neighboring cells influence one another's calcium dynamics at the molecular level(1, 5, 36). Scaling the model to larger cell populations and more complex tissue geometries would help assess whether these dynamics hold at physiologically relevant scales and could open the door to tissue-level predictions of calcium signaling behavior in both healthy and diseased states.

## Conclusion

Our findings show that an $IP_3$-dependent oscillator model with local coupling is sufficient to reproduce the long-range calcium communication patterns observed experimentally. As its behavior depends only on local $IP_3$ concentration and nearest-neighbor coupling, and is governed largely by the balance between coupling and heterogeneity, it can be applied to different cell types, $IP_3$ concentrations, and tissue configurations without changing the underlying architecture. More broadly, these results offer insight into how changes in mechanical state can reorganize tissue-level signaling and provide a foundation for studying the early stages of disease progression in asthma and fibrosis.

## Author contributions

H.P. conceived the research and developed the overall hypothesis. B.M.G. developed the theory, figures and computational model and performed all simulations and analysis. B.M.G. and V.C. drafted the manuscript and edited the figures. B.M.G., V.C., and H.P edited and revised the manuscript. H.P. approved the final version of the manuscript.

## Declaration of interests

The authors declare no competing interests.

## Data Availability

The MATLAB code used to generate all simulations and figures in this study will be deposited in a public repository upon acceptance. Supplementary movies are included with this submission. Any additional information is available from the corresponding author upon reasonable request.

## Acknowledgments

This work was supported by the National Science Foundation (NSF) CAREER grant #2047207 awarded to Harikrishnan Parameswaran.

A preliminary version of this work, arXiv:2607.21442, was deposited on arXiv on July 23, 2026.

## Declaration of generative AI and AI-assisted technologies in the writing process

During the final preparation stages of this manuscript, the authors used Claude (Anthropic) to assist with grammar correction and language editing to improve the clarity and readability of the manuscript. The tool was not used for study design, code development, data generation, analysis, or interpretation. The authors reviewed the output and take full responsibility for the publication.

## References

1. Berridge, M.J., P. Lipp, and M.D. Bootman. 2000. The versatility and universality of calcium signalling. *Nat. Rev. Mol. Cell Biol.* 1:11–21, doi: 10.1038/35036035.
2. Wakai, T., V. Vanderheyden, and R.A. Fissore. 2011. Ca2+ Signaling During Mammalian Fertilization: Requirements, Players, and Adaptations. *Cold Spring Harb. Perspect. Biol.* 3:a006767, doi: 10.1101/cshperspect.a006767.
3. Trexler, A.J., and J.W. Taraska. 2017. Regulation of insulin exocytosis by calcium-dependent protein kinase C in beta cells. *Cell Calcium* 67:1–10, doi: 10.1016/j.ceca.2017.07.008.

4. Klec, C., G. Ziomek, M. Pichler, R. Malli, and W.F. Graier. 2019. Calcium Signaling in ß-cell Physiology and Pathology: A Revisit. *Int. J. Mol. Sci.* 20:6110, doi: 10.3390/ijms20246110.
5. Berridge, M.J. 1993. Inositol trisphosphate and calcium signalling. *Nature* 361:315–325, doi: 10.1038/361315a0.
6. Collier, M.L., G. Ji, Y.-X. Wang, and M.I. Kotlikoff. 2000. Calcium-Induced Calcium Release in Smooth Muscle. *J. Gen. Physiol.* 115:653–662, doi: 10.1085/jgp.115.5.653.
7. Allbritton, N.L., T. Meyer, and L. Stryer. 1992. Range of Messenger Action of Calcium Ion and Inositol 1,4,5-Trisphosphate. *Science* 258:1812–1815.
8. Leybaert, L., and M.J. Sanderson. 2012. Intercellular Ca2+ Waves: Mechanisms and Function. *Physiol. Rev.* 92:1359–1392, doi: 10.1152/physrev.00029.2011.
9. Boitano, S., E.R. Dirksen, and M.J. Sanderson. 1992. Intercellular Propagation of Calcium Waves Mediated by Inositol Trisphosphate. *Science* 258:292–295, doi: 10.1126/science.1411526.
10. National Center for Biotechnology Information. 2004. PubChem Compound Summary for CID 439456, Ins(1,4,5)P3. 2004. https://pubchem.ncbi.nlm.nih.gov/compound/Ins_1_4_5_P3
11. National Center for Biotechnology Information. 2004. PubChem Compound Summary for CID 5460341, Calcium. 2004. https://pubchem.ncbi.nlm.nih.gov/compound/Calcium
12. Clapham, D.E. 1995. Calcium signaling. *Cell* 80:259–268, doi: 10.1016/0092-8674(95)90408-5.
13. Dickinson, G.D., K.L. Ellefsen, S.P. Dawson, J.E. Pearson, and I. Parker. 2016. Hindered cytoplasmic diffusion of inositol trisphosphate restricts its cellular range of action. *Sci. Signal.* 9, doi: 10.1126/scisignal.aag1625.
14. Leybaert, L. 2016. IP3, still on the move but now in the slow lane. *Sci. Signal.* 9:fs17–fs17, doi: 10.1126/scisignal.aal1929.
15. Stasiak, S.E., R.R. Jamieson, J. Bouffard, E.J. Cram, and H. Parameswaran. 2020. Intercellular communication controls agonist-induced calcium oscillations independently of gap junctions in smooth muscle cells. *Sci. Adv.*
16. Kuramoto, Y. 1975. Self-entrainment of a population of coupled non-linear oscillators. In International Symposium on Mathematical Problems in Theoretical Physics. Araki H, editor. Springer-Verlag: Berlin/Heidelberg, pp. 420–422.
17. Strogatz, S.H. 2000. From Kuramoto to Crawford: exploring the onset of synchronization in populations of coupled oscillators. *Phys. Nonlinear Phenom.* 143:1–20, doi: 10.1016/S0167-2789(00)00094-4.
18. Mataragka, S., and C.W. Taylor. 2018. All three IP3 receptor subtypes generate Ca2+ puffs, the universal building blocks of IP3-evoked Ca2+ signals. *J. Cell Sci.* 131:jcs220848, doi: 10.1242/jcs.220848.
19. Bootman, M.D., M.J. Berridge, and P. Lipp. 1997. Cooking with Calcium: The Recipes for Composing Global Signals from Elementary Events. *Cell* 91:367–373, doi: 10.1016/S0092-8674(00)80420-1.
20. Leybaert, L., and M.J. Sanderson. 2012. Intercellular Ca2+ Waves: Mechanisms and Function. *Physiol. Rev.* 92:1359–1392, doi: 10.1152/physrev.00029.2011.
21. Adkins, C.E., and C.W. Taylor. 1999. Lateral inhibition of inositol 1,4,5-trisphosphate receptors by cytosolic Ca2+. *Curr. Biol.* 9:1115–1118, doi: 10.1016/S0960-9822(99)80481-3.

22. Marchant, J.S., and C.W. Taylor. 1997. Cooperative activation of IP3 receptors by sequential binding of IP3 and Ca2+ safeguards against spontaneous activity. *Curr. Biol.* 7:510–518, doi: 10.1016/S0960-9822(06)00222-3.
23. Baker, M.R., G. Fan, V. Arige, D.I. Yule, and I.I. Serysheva. 2023. Understanding IP3R channels: From structural underpinnings to ligand-dependent conformational landscape. *Cell Calcium* 114:102770, doi: 10.1016/j.ceca.2023.102770.
24. Alzayady, K.J., L. Wang, R. Chandrasekhar, L.E. Wagner, F. Van Petegem, and D.I. Yule. 2016. Defining the Stoichiometry of Inositol 1,4,5-Trisphosphate Binding Required to Initiate Ca2+ Release. *Sci. Signal.* 9:ra35, doi: 10.1126/scisignal.aad6281.
25. Schmitz, E.A., H. Takahashi, and E. Karakas. 2022. Structural basis for activation and gating of IP3 receptors. *Nat. Commun.* 13:1408, doi: 10.1038/s41467-022-29073-2.
26. Endo, M. 2009. Calcium-Induced Calcium Release in Skeletal Muscle. *Physiol. Rev.* 89:1153–1176, doi: 10.1152/physrev.00040.2008.
27. Sandler, V.M., and J.-G. Barbara. 1999. Calcium-Induced Calcium Release Contributes to Action Potential-Evoked Calcium Transients in Hippocampal CA1 Pyramidal Neurons. *J. Neurosci.* 19:4325–4336, doi: 10.1523/JNEUROSCI.19-11-04325.1999.
28. Perez, J.F., and M.J. Sanderson. 2005. The Frequency of Calcium Oscillations Induced by 5-HT, ACH, and KCl Determine the Contraction of Smooth Muscle Cells of Intrapulmonary Bronchioles. *J. Gen. Physiol.* 125:535–553, doi: 10.1085/jgp.200409216.
29. Taylor, C.W., and P. Thorn. 2001. Calcium signalling: IP3 rises again… and again. *Curr. Biol.* 11:R352–R355, doi: 10.1016/S0960-9822(01)00192-0.
30. Arenas, A., A. Díaz-Guilera, J. Kurths, Y. Moreno, and C. Zhou. 2008. Synchronization in complex networks. *Phys. Rep.* 469:93–153, doi: 10.1016/j.physrep.2008.09.002.
31. Streb, H., R.F. Irvine, M.J. Berridge, and I. Schulz. 1983. Release of Ca2+ from a nonmitochondrial intracellular store in pancreatic acinar cells by inositol-1,4,5-trisphosphate. *Nature* 306:67–69, doi: 10.1038/306067a0.
32. Wakui, M., B.V.L. Potter, and O.H. Petersen. 1989. Pulsatile intracellular calcium release does not depend on fluctuations in inositol trisphosphate concentration. *Nature* 339:317–320, doi: 10.1038/339317a0.
33. Ma, X., W. Li, and N.L. Stephens. 1997. Heterogeneity of airway smooth muscle at tissue and cellular levels. *Can. J. Physiol. Pharmacol.* 75:930–935.
34. Hirst, S.J. 2006. SMOOTH MUSCLE CELLS | Airway. In Encyclopedia of Respiratory Medicine. Elsevier, pp. 96–105.
35. Radicchi, F., and H. Meyer-Ortmanns. 2006. Entrainment of coupled oscillators on regular networks by pacemakers. *Phys. Rev. E* 73:036218, doi: 10.1103/PhysRevE.73.036218.
36. Chandar, V., B.M. Goykadosh, and H. Parameswaran. 2026. Intercellular contact is sufficient to drive fibroblast-to-myofibroblast transitions. *J. Appl. Physiol.* 140:1592–1604, doi: 10.1152/japplphysiol.00382.2025.